\documentclass[a4paper,11pt]{article}
\usepackage{pos}
\def\qon#1{q_{#1,0}^{(+)}}
\def\Eq#1{Eq.~(\ref{#1})}

\title{From vacuum amplitudes to qubits}

\author*[a]{Germ\'an Rodrigo}

\affiliation[a]{Instituto de F\'{\i}sica Corpuscular, Universitat de Val\`{e}ncia -- Consejo Superior de Investigaciones Cient\'{\i}ficas, Parc Cient\'{\i}fic, E-46980 Paterna, Valencia, Spain}

\emailAdd{german.rodrigo@csic.es}

\abstract{High-energy colliders, exemplified by the CERN's Large Hadron Collider (LHC), constitute genuine quantum machines. In alignment with Richard Feynman's foundational vision for quantum computing, collider physics emerge therefore as a prime candidate for quantum simulations. Prospective applications include Quantum Machine Learning for collider data analysis, accelerated evaluation of complex multiloop Feynman diagrams, efficient jet clustering, enhanced parton shower simulations, and related computational challenges. We discuss two specific applications: the identification of causal structures in multiloop vacuum amplitudes, a fundamental component of the Loop-Tree Duality exhibiting deep connections to graph theory; and high-dimensional function integration and sampling. The latter constitutes an initial step toward realizing a fully fleged quantum event generator capable of operating at high perturbative orders.}

\FullConference{17th International Symposium on Radiative Corrections: Applications of Quantum Field Theory to Phenomenology (RADCOR2025)\\
5-10 October 2025\\
Puri, India\\}

\begin{document}
\maketitle


\section{Introduction}

High-energy physics is entering an era of unprecedented experimental precision. The upcoming High-Luminosity phase of the CERN's Large Hadron Collider (HL-LHC) will enable extraordinarily accurate measurements of fundamental properties of elementary particles, such as most Higgs boson couplings to percent-level accuracy. This experimental achievement imposes a strategic imperative on the theoretical physics community to advance predictions at a commensurate pace. The precision gap between theory and experiment is not a remote prospect, but an imminent challenge. Projections from ATLAS and CMS~\cite{Cepeda:2019klc} indicate that theoretical predictions will dominate the total uncertainties for key HL-LHC observables. This mismatch, in which experimental performance surpasses theory, is unsustainable for making solid claims about possible discoveries and demands a paradigm shift in computational methods.

Quantum Field Theory (QFT) is the cornerstone of our understanding of physics at high-energy colliders, as the interactions between elementary particles are inherently Quantum Mechanical (QM). This fact establishes colliders as genuine quantum machines, in particular, with {\bf the LHC being the largest quantum machine ever constructed}. Quantum computing (QC) thus offers a natural framework for novel quantum methodologies in collider physics, where many tasks promise to be far more efficient than classical approaches. This novel direction is also in accordance with the seminal insight of Richard P. Feynman for QC: {\it Nature isn’t classical, dammit, and if you want to simulate it, your simulator had better be quantum}.

The central computational challenge in QFT predictions lies in calculating scattering amplitudes. These are systematically organised via Feynman diagrams, which provide a perturbative expansion encompassing all possible particle interaction pathways. However, as scattering amplitudes are defined for a fixed number of external particles, they represent nonphysical objects because the number of external particles is not a QM observable, particularly in gauge theories such as Quantum Chromodynamics (QCD), where gluons can be emitted at exactly zero energy, for example. In standard QFT calculations, particle emission at zero energy is mathematically distinguished from the absence of emission, introducing an inconsistency which is also manifest in the artificial separation of loop and tree-level contributions. For this reason, Refs.~\cite{Ramirez-Uribe:2024rjg, LTD:2024yrb} have introduced the idea of basing QFT predictions on vacuum amplitudes, i.e. scattering amplitudes that do not involve external particles. This approach provides important conceptual and technical advantages, which are rooted to the manifest causal properties of the Loop-Tree Duality (LTD)~\cite{Aguilera-Verdugo:2020set,Aguilera-Verdugo:2020kzc,Ramirez-Uribe:2020hes} (See also Ref.~\cite{Kermanschah:2026odl,Kermanschah:2026dzb} in this conference).

\begin{figure}[t]
\begin{center}
\includegraphics[width=.9\textwidth]{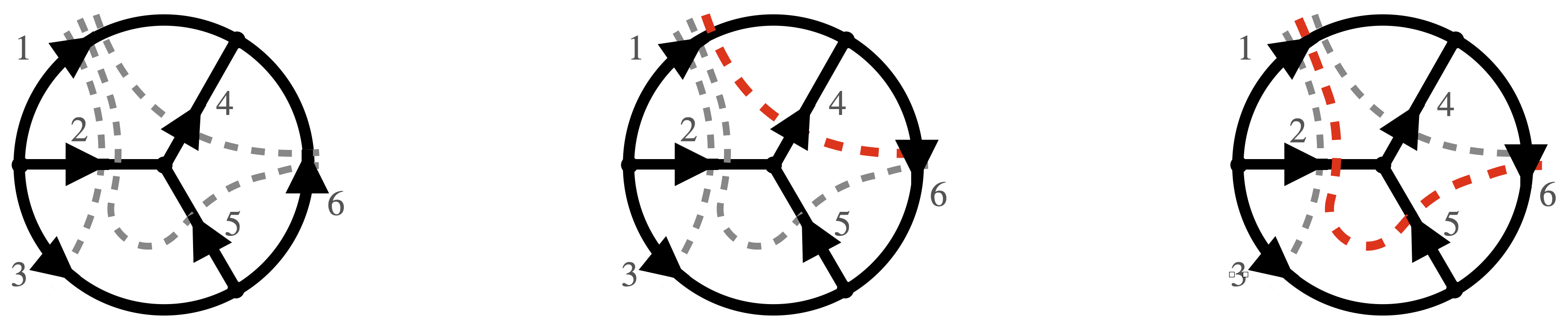}
\end{center}
\begin{equation}
{\cal A}_{\rm D}^{(\Lambda)} \sim \frac{\displaystyle 1}{\displaystyle \lambda_{123} \lambda_{1256} \lambda_{146}} \qquad
{\rm Res} \left({\cal A}_{\rm D}^{(\Lambda)},\lambda_{146} \right)  \sim \frac{1}{\lambda_{123} \lambda_{25\bar 4}} \qquad 
{\rm Res} \left({\cal A}_{\rm D}^{(\Lambda)},\lambda_{1256} \right)  \sim \frac{-1}{\lambda_{123} \lambda_{25\bar 4}} \nonumber
\end{equation}
\caption{Three-loop vacuum amplitude (left). Phase-space residue with three external particles representing the interference of a one-loop amplitude with a tree-level amplitude (centre). Phase-space residue with four external particles representing the interference of two tree-level amplitudes (right).}
\label{fig:vacuum}
\end{figure}

In LTD, after integrating out the energy components of the $\Lambda$ primitive loop four-momenta, $\{\ell_s\}_{s=1}^\Lambda$, scattering amplitudes are expressed as functions of on-shell energies, $\qon{i_s} = \sqrt{{\bf q}_{i_s}^2 + m_{i_s}^2 - \imath 0}$. Specifically, vacuum amplitudes are written in LTD in terms of sums of on-shell energies, $\lambda_{i_1\cdots i_n} = \sum_{s=1}^n \qon{i_s}$. In this representation, the physical interpretation is particularly insightful. Each $\lambda_{i_1\cdots i_n}$ corresponds to a set of propagators that partition the vacuum amplitude into two subamplitudes, with all propagators in that set sharing a consistent direction of propagation. By analytically continuing the on-shell energies of the particles that will be identified as incoming to negative values, final states are generated from residues, called phase-space residues, over specific $\lambda_{i_1\cdots i_n}$ on the vacuum amplitude. An illustrative three-loop example is presented in Fig.~\ref{fig:vacuum}, where $\lambda_{25\bar 4} = \lambda_{25} - \lambda_{4}$ encodes the collinear splitting $4\to 25$. Relevant is the change of sign from one phase-space residue to the other, which guaranties a local cancellation of soft and collinear singularities. This local cancellation of singularities at the integrand level occurs because loop and tree-level contributions are implemented under a common integration domain. For a decay rate 
\begin{equation}
d\Gamma_{{\rm N}^k{\rm LO}}
\sim \int_{\vec{\ell}_1 \cdots \vec{\ell}_{\Lambda-1}} \sum {\rm Res} \left( {\cal A}_{\rm D}^{(\Lambda, {\rm R})}, \lambda_{i_1 i_2\cdots i_n a} \right) \delta\left( \lambda_{i_1 i_2 \dots i_n} - \lambda_{a}\right)~,
\label{eq:decay}
\end{equation}
where the label $a$ represents the decaying particle, and $\delta\left( \lambda_{i_1 i_2 \dots i_n} - \lambda_{a}\right)$ corresponds to energy conservation. The superindex ${\rm R}$ indicates that the phase-space residue is renormalised in the ultraviolet (UV) regime by suitable local counterterms. For hadronic collisions, these counterterms also locally subtract initial-state collinear singularities.  

The expression in \Eq{eq:decay} represents a complex, high-dimensional integral over the three-momenta of all possible intermediate (virtual) particle states and the phase space of final-state particles. The High-Luminosity LHC's precision requirements demand calculations involving multiloop Feynman diagrams with numerous external legs, which exponentially increase both the computational complexity and the dimensionality of these integrals. At Next-to-Leading Order~(NLO), a $1\to 2$ decay requires two independent integration variables, whereas a $2\to 2$ scattering in $e^+e-$ annihilation needs four. For hadronic collisions, integration over the two longitudinal momentum fractions from  parton densities should also be considered. Each additional order  requires three extra dimensions. At Next-to-Next-to-Leading Order (NNLO) in hadron colliders, $2\to 2$ scattering involves nine integration variables, while $2\to 3$ scattering requires twelve.

\section{A new sort of Qubits motivated by Causality}

Causality constitutes a fundamental principle of physics requiring physical processes to respect temporal ordering such that effects cannot precede their causes. Within Feynman diagrams, this requirement imposes a strict constraint on momentum flow and propagation paths, prohibiting closed cycles. Physically admissible configurations manifest as Directed Acyclic Graphs (DAGs) in a graph-theory perspective, whereas cyclic configurations are unphysical, as they entail the momentum flow returning to the emission point and thus traveling back in time against causality.

The Feynman representation of scattering amplitudes need not specify particle propagation directions explicitly, as its manifest Lorentz covariance encodes all time orderings within a single compact expression. However, the inclusion of cyclic configurations introduces spurious singularities in the integrand.  LTD addresses this inconsistency by being manifestly causal. Vacuum amplitudes in LTD retain only acyclic, physical configurations yielding less singular integrands that are more suitable to numerical integration.

Motivated by the necessity to identify acyclic configurations within LTD, we define a {\bf new sort of qubit}, the Feynman propagator. A Feynman propagator encodes the quantum superposition of a particle propagating between two interaction vertices in either direction, thereby constituting a genuine two-state quantum system. In QM notation, the following analogy naturally arises:
\begin{equation}
G_{\rm F} (q_i) = \frac{1}{q_i^2 - m_i^2 + \imath 0} \equiv \frac{1}{\sqrt{2}} \left( |0\rangle  + |1\rangle\right)~,
\end{equation}
where $|0\rangle$ denotes propagation in the forward direction and $|1\rangle$ in the reverse direction. A Feynman diagram, or scattering amplitude, thus corresponds to a quantum superposition of $2^n$ states, of which only a subset are acyclic and thus physically meaningful. Isolating these acyclic states, which represent causal configurations, makes causality manifest in theoretical calculations. Conversely, by fixing the propagation directions, the integrand representation in~LTD can be bootstrapped.

\section{Mapping Cycles to Multicontrolled Toffoli Quantum Gates}

The task of identifying the subset of causal DAGs within the full set of possible momentum-flow configurations in multiloop Feynman diagrams can be recast as an unstructured search problem over an exponentially large configuration space. This formulation naturally lends itself to QC, particularly to search algorithms based on Grover’s framework~\cite{Grover:1997fa,Ramirez-Uribe:2021ubp}. Alternatively, the problem can be expressed as an energy minimization task, where the objective is to determine the ground state of a Hamiltonian encoding the number of cycles in a graph. From this perspective, the Variational Quantum Eigensolver~(VQE) approach becomes suitable~\cite{Clemente:2022nll}.

Given the current limitations of quantum hardware, the primary objective is not necessarily to achieve a significant computational speedup in identifying the physically relevant solutions, but rather to develop insight into the design and optimization of the quantum oracle. The general workflow of a Grover-based algorithm, along with its corresponding quantum circuit, requires three steeps. The algorithm employs a central set of qubits, referred to as the edge register, whose size equals the number of internal Feynman propagators. These qubits are initialized in a uniform superposition of $|0\rangle$ and $|1\rangle$ states, typically prepared by applying Hadamard gates to each qubit. In this representation, the state $|0\rangle$ for a given qubit corresponds to particle propagation between two interaction vertices in one predefined direction, whereas the state $|1\rangle$ denotes propagation in the opposite direction.

In the second step, the oracle marks those entangled states in the edge register that correspond to valid DAG configurations by applying a conditional phase flip, however, leaving the underlying probability distribution of the qubits unchanged. Implementing this tagging procedure may require additional ancillary qubits to store intermediate results or to perform arithmetic and logical operations on the edge assignments. Once the oracle has marked the causal states, the final step consists of a diffusion operator that performs a rotation in the Hilbert space to amplify the probability amplitudes of the marked states while suppressing those of the unmarked ones. As a result, measuring the qubits in the edge register yields a valid DAG configuration with higher probability. Repeating the procedure iteratively allows the identification of all valid causal DAGs.

A key innovation was introduced in Ref.~\cite{Ramirez-Uribe:2024wua} in the design of the quantum oracle, where cycles are detected using multicontrolled Toffoli gates. Such a gate operates on multiple control qubits and flips the state of a target qubit via an $X$ gate only when all control qubits are in the $|1\rangle$ state; otherwise, the target qubit remains unchanged. Since cycles in a Feynman diagram correspond to configurations where all propagators share the same orientation, the multicontrolled Toffoli gate serves as the natural quantum analogue. Beyond enriching the conceptual correspondence between QC and Feynman diagrams, this oracle construction substantially reduces the implementation cost for specific classes of diagrams, leading to remarkable improvements in practical runtime on quantum simulators and potentially on quantum hardware.

\section{Graph-Theory Principles to Optimize the Quantum Oracle}

\begin{figure}[t]
\begin{center}
    \includegraphics[scale = 0.38]{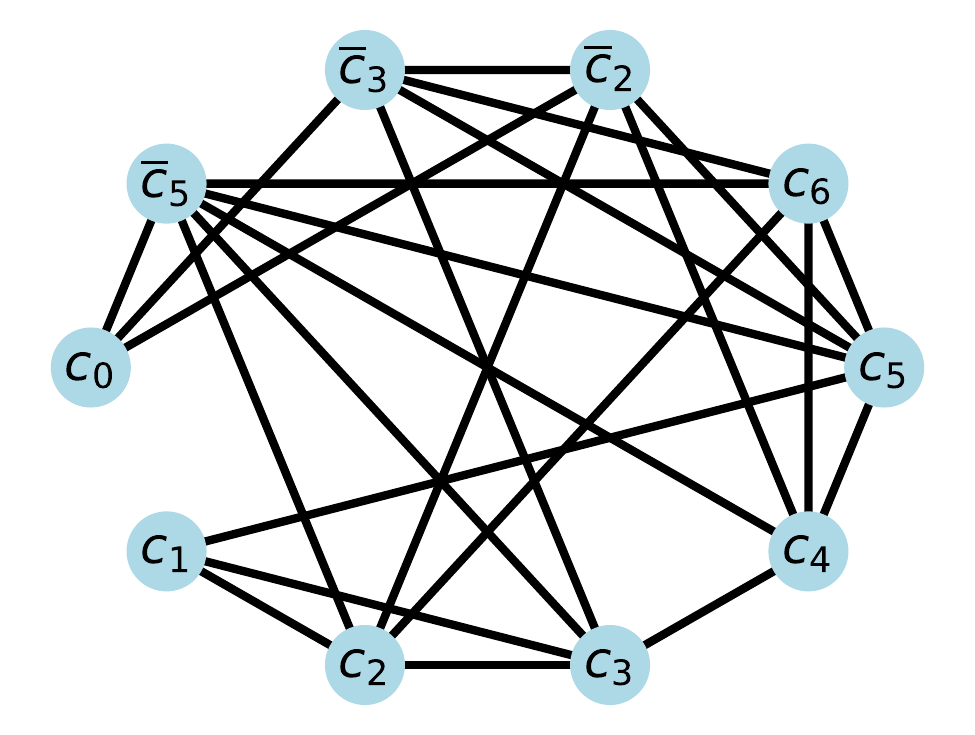}
    \includegraphics[scale = 0.40]{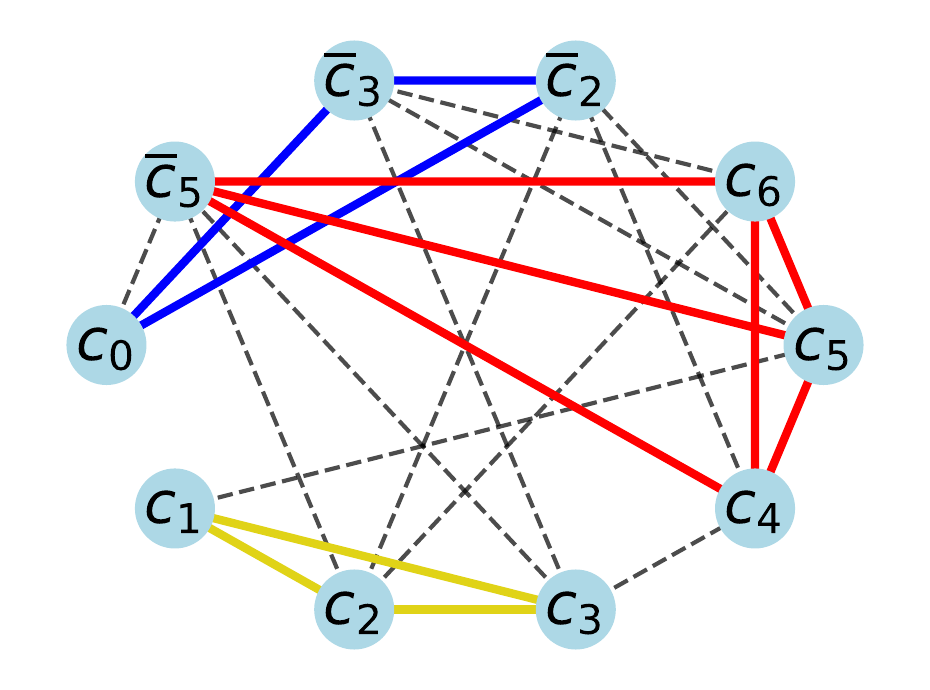}
\end{center}    
    \caption{
    Graph representing the adjacency matrix of mutually exclusive clauses of a three-eloop topology with twelve edges, generated with the \texttt{MutualAuxMatrix} algorithm (left), and corresponding \texttt{MAUXc}$^{(3,12)}$ obtained with the \texttt{GraphConditionCombination} algorithm (right)~\cite{Ochoa-Oregon:2025opz}. Colours indicate the different cliques.}
    \label{fig:graph_mutual_exclusive_2}   
\end{figure}

The efficiency of a Grover-based search algorithm depends critically on the resource demands of the quantum oracle. Given the deep ties to graph theory, it seems natural to leverage fundamental principles from this field to improve the design of the oracle. The key idea presented in Ref.~\cite{Ochoa-Oregon:2025opz} is to analyse the logical relationships between cycles implemented using multicontrolled Toffoli gates and determine which ones are mutually exclusive.  From these, one constructs the adjacency graph of Mutually Exclusive Clauses (MEC). This transforms the oracle optimisation into a Minimum Clique Partition (MCP) problem, consisting of identifying the smallest collection of fully connected subgraphs (cliques) required to cover all the vertices in the MEC graph.

This optimisation substantially reduces the number of ancillary qubits required. Since all clauses within a single clique are mutually exclusive, the information they contain can be encoded using a single ancillary qubit rather than one per clause. See Fig.~\ref{fig:graph_mutual_exclusive_2} for an illustrative three-loop example. The graph on the left represents the adjacency matrix of MECs. It is clear that $c_i$ and $\bar c_i$ are MECs; the rest of the relations are less evident and require the performance of some Boolean algebra. From Fig.~\ref{fig:graph_mutual_exclusive_2} (right), three cliques can be identified. Thus, this approach reduces the ancillary qubit count from seven to just three. This reduction in qubit overhead is a significant advancement that could enable the analysis of physically relevant multiloop diagrams on near-term Noise Intermediate Scale Quantum (NISQ) devices.

\section{Quantum Integration of Multidimensional Functions}

The accuracy of theoretical predictions at particle colliders depends on accurately accounting for loop quantum fluctuations at high perturbative orders. Each additional loop introduces extra dimensions to the loop integrals, while each radiated particle adds further dimensions to the phase space. In this context, QC offer a potential advantage, since the computational complexity grows exponentially with dimensionality. Moreover, QC provides the most natural framework for high-dimensional sampling, or event generation in the language of high-energy physics. We focus on LTD, which provides a manifestly causal representation that yields numerically stable integrands and unifies loop and phase-space integration. Two novel quantum algorithms have been developed that specifically target this challenge, QFIAE~\cite{deLejarza:2023qxk} and QAIS~\cite{Pyretzidis:2025stx,Pyretzidis:2025txb}. Other quantum integration methods include~\cite{Yi:2025feo,Williams:2025hza,Cruz-Martinez:2023vgs,Agliardi:2022ghn}.

\subsection{Quantum Fourier Iterative Amplitude Estimation (QFIAE)}

\begin{figure}[t]
\begin{center}
\begin{tabular}{c}
\includegraphics[scale=.35]{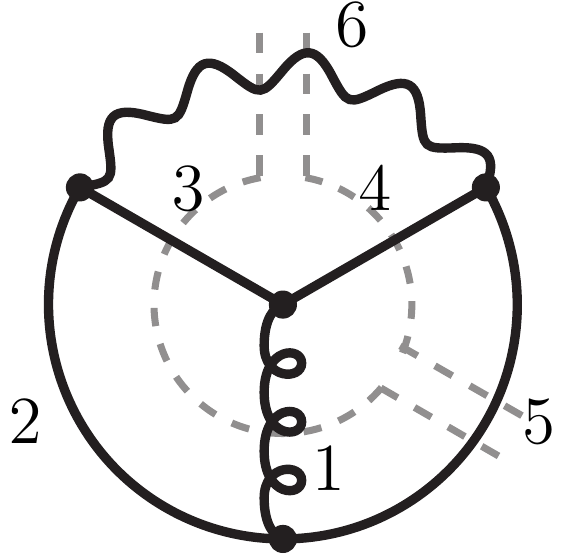} 
\includegraphics[scale=.35]{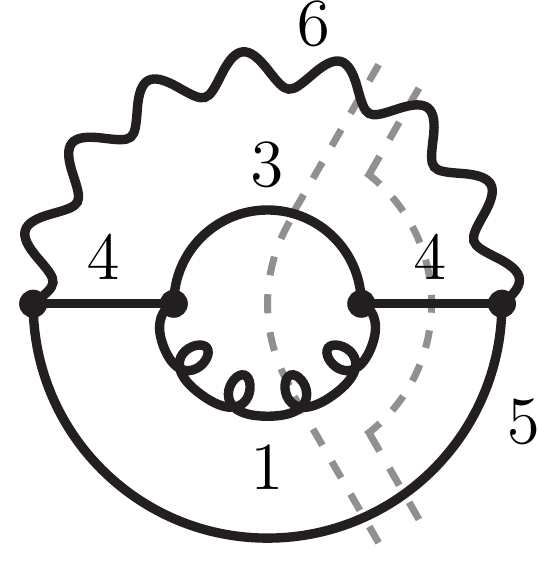} 
\end{tabular}
\begin{tabular}{c}
\includegraphics[scale=.5]{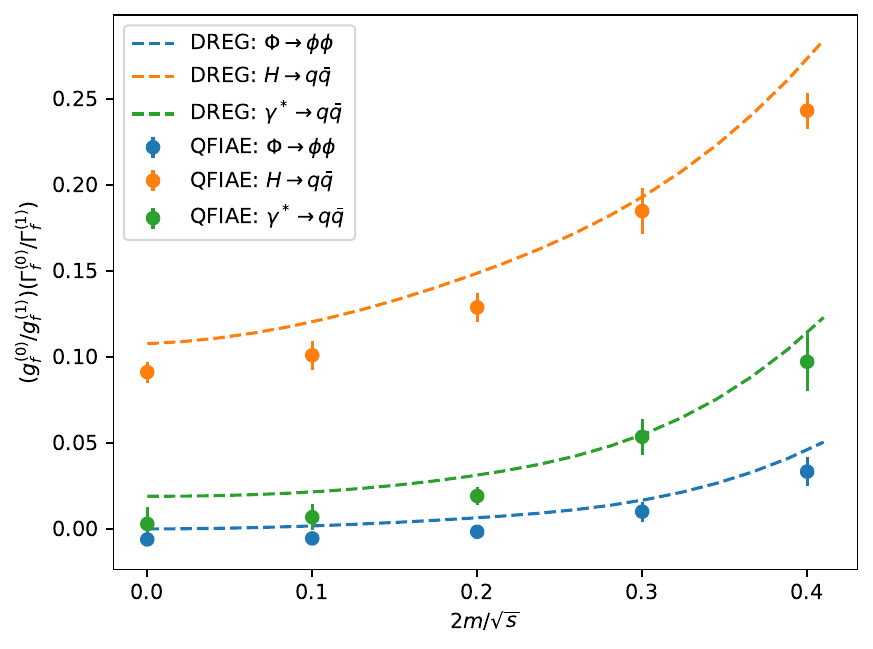}
\end{tabular}
\caption{Vacuum diagrams contributing to the decay rate $\gamma^*\to q\bar q (g)$ at NLO (left), and integrated results as a function of the quark mass using QFIAE partially in quantum hardware (right). 
\label{fig:lhc_integral}}
\end{center}
\end{figure}

The QFIAE~\cite{deLejarza:2023qxk} method is a hybrid approach that combines Quantum Machine Learning~(QML) and Quantum Amplitude Estimation~(QAE). The procedure comprises two steps: first, a Quantum Neural Network (QNN) is trained to derive an accurate and compact Fourier series representation of the target integrand, effectively decomposing a potentially complex function into a sum of simpler trigonometric components. Second, the Iterative Quantum Amplitude Estimation~(IQAE)~\cite{Grinko:2021iad} algorithm, a variant of Grover's algorithm, is used to integrate each trigonometric component of the Fourier series. The final integral is obtained by summing these contributions.

This method has been successfully applied to computing Feynman integrals in LTD~\cite{deLejarza:2024pgk} and decay rates at Next-to-Leading Order (NLO) for several physical processes~\cite{deLejarza:2024scm}, employing vacuum amplitudes. The close agreement between the results obtained using a quantum simulator and the established DREG method validates the QFIAE approach. While results on actual quantum hardware (Fig.~\ref{fig:lhc_integral}) demonstrate the impact of current device noise, they nevertheless confirm the viability of the algorithm on real quantum processors.

\subsection{Quantum Adaptive Importance Sampling (QAIS)}

\begin{figure}[t]
\begin{center}
\includegraphics[width=.8\textwidth]{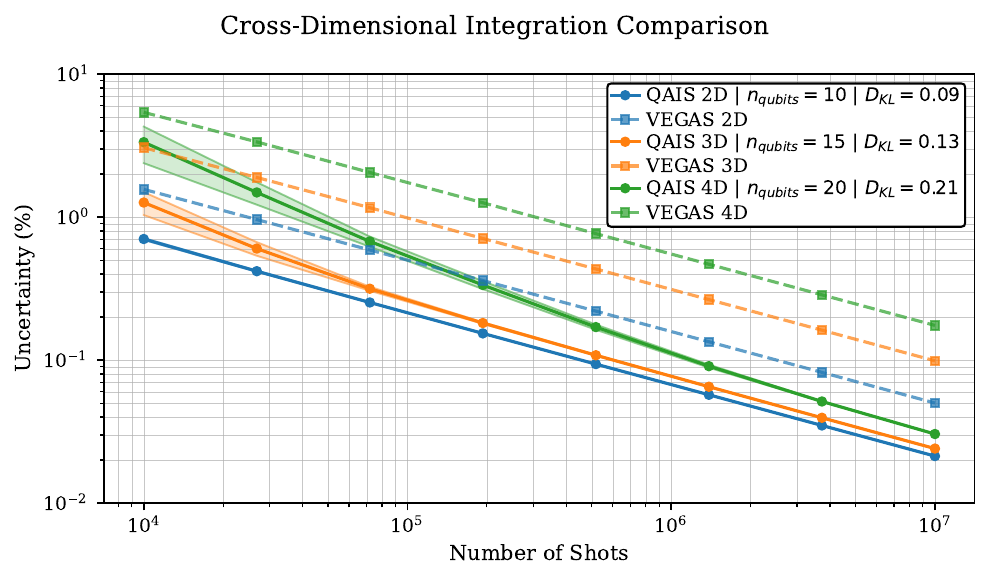}
\caption{Average uncertainty for 100 independent integration runs using the same proposal PDF for the benchmark integral in~\Eq{eq:benchmark_int}, across the full standard MC sampling range. The shaded areas, correspond to the $\pm 1 \sigma$ dispersion of these 100 runs. The VEGAS result presented corresponds to its best iteration.  }
\label{fig:4Dminimal}
\end{center}
\end{figure}

Classical Monte Carlo integration commonly employs Adaptive Importance Sampling to improve efficiency, with VEGAS~\cite{Lepage:1977sw,Lepage:2020tgj,Hahn:2004fe} as the most prominent and widely used implementation. However, the complexity of integration grids increases exponentially with dimensionality, which limits the effectiveness of these algorithms and compels the use of separable grids. Consequently, integrands with correlated structures whose relevant regions are not aligned with the coordinate axes frequently result in phantom peaks, i.e. oversampled regions that have little contribution.

The Quantum Adaptive Importance Sampling (QAIS) algorithm~\cite{Pyretzidis:2025stx,Pyretzidis:2025txb} is specifically designed to overcome this limitation. It employs a Parametrized Quantum Circuit (PQC) to construct a non-separable proposal Probability Density Function (PDF), which is trained to closely approximate the target integrand, regardless of its complexity. In this way, QAIS directs PQC shots towards the most important regions of the integration domain, thereby maximising sampling efficiency and substantially reducing the number of function evaluations needed to achieve a specified level of precision. It also accommodates correlated structures that separable grids cannot capture. Furthermore, QAIS incorporates a tiling mechanism that compensates for the bias caused by a finite number of shots, systematically covering missing regions to ensure an unbiased estimator while maintaining the efficiency gains from the learned PDF.

The good scaling of QAIS relative to VEGAS has been demonstrated by comparing their performance on a multipeak benchmark integral across increasing dimensions, and with a pentagon Feynman integral at one-loop, which is a three-dimensional integral in LTD. For example, a performance comparison for the function 
\begin{equation}\label{eq:benchmark_int}
\int_{[0,1]^d} f(\mathbf{x}) d \mathbf{x} = \int_{[0,1]^d} \left( \sum_{i=0}^{2} e^{-50 |\mathbf{x} - \mathbf{r}_i |} \right) d \mathbf{x}
\end{equation}
where $\mathbf{r_0}=(0.23,\dots,0.23)$, $\mathbf{r_1}=(0.39,\dots,0.39)$, and $\mathbf{r_2}=(0.74,\dots,0.74)$, is presented in Fig.~\ref{fig:4Dminimal}. We consider the cases $d=2,3,4$, using the same discretisation in all instances, which is five qubits per dimension. QAIS consistently delivers a lower uncertainty for the same number of shots, with the performance gap widening as dimensionality increases. The PQC requires 225-750 parameters. Preliminary comparisons with Classical Machine Learning (ML) based integrators~\cite{Heimel:2022wyj,Winterhalder:2021ngy,Bothmann:2020ywa,Gao:2020vdv} indicates that QAIS requires substantially fewer trainable parameters to properly capture correlations.

\section{Conclusions}

High-energy colliders are genuine quantum machines, in which particle interactions are governed by the laws of QFT. This intrinsic quantum nature makes colliders a natural and compelling domain for the design and benchmarking of QC algorithms. Causality, a fundamental principle in physics, determines in the context of Feynman diagrams the physically valid propagation directions of virtual particles. In graph-theory terms, physical configurations correspond to DAGs. A conceptual mapping emerges between particle physics and QC, whereby Feynman propagators are qubits and cycles are encoded by multicontrolled Toffoli gates. This analogy facilitates the development of new QC algorithms and provides a robust framework for optimising quantum oracle design through graph-theory concepts. We have developed and validated two quantum integration algorithms, QFIAE and QAIS, both of which are powerful and viable tools for tackling multidimensional integrals, which are prevalent in high-energy physics. Notably, QAIS distinguishes itself through its ability to capture highly correlated integrands and its advantageous scaling with dimensionality, making it a particularly promising solution for the increasing computational demands of future calculations.

\section*{acknowledgments}
This work is supported by the Spanish Government and ERDF/EU - Agencia Estatal de Investigaci\'on MCIN/AEI/10.13039/501100011033,  Grants No. PID2023-146220NB-I00, No. EUR2025-164820, and No. CEX2023-001292-S; Generalitat Valenciana, Grant No. ASFAE/2022/009 (Planes Complementarios de I+D+i, NextGenerationEU); and Quantum Spain (Ministry of Economic Affairs and Digital Transformation,  NextGenerationEU).

\bibliographystyle{JHEP}
\bibliography{2025_RADCOR_rodrigo}

\end{document}